\documentclass[11pt,epsf,epsfig]{article}
\usepackage{graphics}
\usepackage{graphicx}
\usepackage{arydshln}						
\usepackage{amssymb}		
\usepackage{amsmath}

\newcommand{\gsim}{\lower.7ex\hbox{$\;\stackrel{\textstyle>}{\sim}\;$}}
\newcommand{\lsim}{\lower.7ex\hbox{$\;\stackrel{\textstyle<}{\sim}\;$}}

\def\stilde{\widetilde}

\newcommand{\newc}{\newcommand}
\newc{\Nc}{N_{c}}
\newc{\CG}{C_G}
\newc{\gp}{g'}
\newc{\stopi}{\stilde t_i}
\newc{\sboti}{\stilde b_i}
\newc{\staui}{\stilde \tau_i}
\newc{\stopj}{\stilde t_j}
\newc{\sbotj}{\stilde b_j}
\newc{\stauj}{\stilde \tau_j}
\newc{\stopI}{\stilde t_1}
\newc{\stopII}{\stilde t_2}
\newc{\sbotI}{\stilde b_1}
\newc{\sbotII}{\stilde b_2}
\newc{\stauI}{\stilde \tau_1}
\newc{\stauII}{\stilde \tau_2}
\newc{\sstop}{s_{t}}
\newc{\cstop}{c_{t}}
\newc{\ssbot}{s_{b}}
\newc{\csbot}{c_{b}}
\newc{\sstau}{s_{\tau}}
\newc{\cstau}{c_{\tau}}
\newc{\Sstop}{s_{2t}}
\newc{\Cstop}{c_{2t}}
\newc{\Ssbot}{s_{2b}}
\newc{\Csbot}{c_{2b}}
\newc{\Sstau}{s_{2\tau}}
\newc{\Cstau}{c_{2\tau}}
\newc{\salpha}{s_\alpha}
\newc{\calpha}{c_\alpha}
\newc{\Calpha}{c_{2\alpha}}
\newc{\Salpha}{s_{2\alpha}}
\newc{\sbetapm}{s_{\beta_\pm}}
\newc{\cbetapm}{c_{\beta_\pm}}
\newc{\Sbetapm}{s_{2 \beta_\pm}}
\newc{\Cbetapm}{c_{2 \beta_\pm}}
\newc{\sbetaO}{s_{\beta_0}}
\newc{\cbetaO}{c_{\beta_0}}
\newc{\SbetaO}{s_{2 \beta_0}}
\newc{\CbetaO}{c_{2 \beta_0}}
\newc{\vu}{v_u}
\newc{\vd}{v_d}
\newc{\seL}{\stilde e_L}
\newc{\smuL}{\stilde \mu_L}
\newc{\seR}{\stilde e_R}
\newc{\smuR}{\stilde \mu_R}
\newc{\suL}{\stilde u_L}
\newc{\sdL}{\stilde d_L}
\newc{\suR}{\stilde u_R}
\newc{\sdR}{\stilde d_R}
\newc{\scL}{\stilde c_L}
\newc{\ssL}{\stilde s_L}
\newc{\scR}{\stilde c_R}
\newc{\ssR}{\stilde s_R}
\newc{\snue}{\stilde \nu_e}
\newc{\snumu}{\stilde \nu_\mu}
\newc{\snutau}{\stilde \nu_\tau}
\newc{\Gpm}{G^\pm}
\newc{\Hpm}{H^\pm}
\newc{\FFbS}{\overline{FF}S}
\newc{\FFbV}{\overline{FF}V}
\newc{\FSS}{F_{SS}}
\newc{\FSSS}{F_{SSS}}
\newc{\FFFS}{F_{FFS}}
\newc{\FFFbS}{F_{\overline{FF}S}}
\newc{\FSSV}{F_{SSV}}
\newc{\FVS}{F_{VS}}
\newc{\FVVS}{F_{VVS}}
\newc{\FFFV}{F_{FFV}}
\newc{\FFFbV}{F_{\overline{FF}V}}
\newc{\Fgauge}{F_{\rm gauge}}
\newc{\DRbarprime}{$\overline{\rm DR}'$ }
\newc{\DRbar}{$\overline{\rm DR}$ }
\newc{\MSbar}{$\overline{\rm MS}$ }
\newc{\Yu}{{\bf Y}_u}
\newc{\Yd}{{\bf Y}_d}
\newc{\Ye}{{\bf Y}_e}
\newc{\Au}{{\bf a}_u}
\newc{\Ad}{{\bf a}_d}
\newc{\Ae}{{\bf a}_e}
\newc{\bm}{{\bf m}}

\newc{\rwino}{r_{\tilde W}}
\newc{\rmu}{r_{\tilde H}}
\newc{\ra}{r_A}

\newcommand{\nc}{\newcommand}

\nc{\beaa}{\begin{eqnarray*}} \nc{\eeaa}{\end{eqnarray*}}
\nc{\beq}{\begin{equation}}   \nc{\eeq}{\end{equation}}
\nc{\bea}{\begin{eqnarray}}   \nc{\eea}{\end{eqnarray}}
\nc{\baa}{\begin{array}}      \nc{\eaa}{\end{array}}
\nc{\bit}{\begin{itemize}}    \nc{\eit}{\end{itemize}}
\nc{\ben}{\begin{enumerate}}  \nc{\een}{\end{enumerate}}
\nc{\bce}{\begin{center}}     \nc{\ece}{\end{center}}
\nc{\non}{\nonumber}

\def\bed{\begin{description}}
\def\eed{\end{description}}

\def\non{\nonumber}

\def\k1slash{k_1\hspace{-10.5pt}/\ \ }

\def\simge{\mathrel{%
   \rlap{\raise .57ex \hbox{$>$}}{\lower .57ex \hbox{$\sim$}}}}
\def\simle{\mathrel{
   \rlap{\raise 0.512ex \hbox{$<$}}{\lower 0.512ex \hbox{$\sim$}}}}

\begin{document}
\setlength{\baselineskip}{0.22in}

\title{\textbf{Lepton Flavor Violation and Supersymmetric}\\
\textbf{Dirac Leptogenesis}}

\author{Brooks Thomas and Manuel Toharia \\
\it{Michigan Center for Theoretical Physics (MCTP)} \\
\it{Department of Physics, University of Michigan, Ann Arbor, MI 48109}}

\date{MCTP 06-17\\ July 25, 2006}
\maketitle

\begin{abstract}
Dirac leptogenesis (or Dirac neutrinogenesis), in which neutrinos are purely
Dirac particles, is an interesting alternative to the
standard leptogenesis scenario.  In its supersymmetric version,
the modified form of the superpotential required for successful
baryogenesis contributes new,
generically non-flavor-diagonal terms to the slepton and sneutrino mass
matrices.
In this work,
we examine how current
experimental bounds on flavor-changing effects in the lepton sector
(and particularly the bound on \(\mu\rightarrow e\gamma\)) constrain
Dirac leptogenesis and we find that it is capable of succeeding 
with superpartner 
masses as low as $\sim 100$~GeV. For such light 
scalars and electroweakinos, upcoming experiments such as MEG are
generically expected to observe signals of lepton flavor violation.
\end{abstract}

\section{Introduction\label{sec:Intro}}

\indent

When singlet fermions are not present in a given theory, Dirac
leptogenesis~\cite{Dick:1999je,Murayama:2002je}, or Dirac
neutrinogenesis, represents a very interesting alternative to the
traditional leptogenesis scenario, which relies on the existence of
heavy Majorana neutrinos. In Dirac leptogenesis, neutrinos are
purely Dirac particles whose small but nonzero masses appear as
ratios of dimensionful parameters in an effective field theory.  It
has been shown~\cite{Thomas:2005rs} that, in the context of split
supersymmetry~\cite{Arkani-Hamed:2004fb,Giudice:2004tc}, Dirac
leptogenesis is a phenomenologically viable scenario capable of
satisfying all relevant constraints from cosmology and neutrino
physics as well as reproducing the observed baryon-to-photon ratio
\(\eta\) of the universe. Split supersymmetry is advantageous
primarily for two reasons. The first of these is that very heavy
gravitinos can easily evade the constraints that big bang
nucleosynthesis (BBN) places on the post-inflationary reheating
temperature; the second is that in Dirac leptogenesis the
superpotential is extended by the addition of new terms carrying
phases and nontrivial flavor structure. Dangerous contributions from
these terms to flavor-changing processes are automatically safe in
split supersymmetry, but must be treated with care when the scale of
all the superpartners is near the electroweak scale.
In the latter case (assuming that the constraints for gravitino
cosmology are satisfied) it is necessary to compute carefully the
rates for flavor-violating processes in the lepton sector.

\indent The aim of this paper is to ascertain whether or not Dirac
Leptogenesis is permitted by flavor violation constraints in
supersymmetric theories with low-scale slepton masses.  We will
begin our inquiry by briefly reviewing the theoretical framework of
Dirac leptogenesis and deriving the additional contributions to the
slepton mass matrices to which its superpotential gives rise.  We
then turn to the calculation of rates for the processes
\(\mu\rightarrow e\gamma\) and \(\tau\rightarrow \mu\gamma\).
Finally, we apply the combined constraints from flavor violation and
baryogenesis and assess the viability and predictability of Dirac
leptogenesis when superpartner masses are at or around the weak
scale.

\section{Dirac Leptogenesis and the Slepton Mass Matrices\label{sec:DiracLep}}

\indent

In Dirac leptogenesis, as in the traditional leptogenesis
picture~\cite{Fukugita:1986hr,Luty:1992un}, the conditions for
successful baryogenesis~\cite{Sakharov:1967dj} are met by positing
the existence of a heavy particle with CP-violating decays into
leptons.\footnote{For recent variations on both
supersymmetric and non-supersymmetric scenarios,
see for example~\cite{Cerdeno:2006ha,Abel:2006hr}.} The lepton number produced
in these decays is then processed into a baryon number for the
universe by electroweak sphaleron processes~\cite{'tHooft:1976fv}.
In the supersymmetric version of Dirac leptogenesis, all of this is
engineered via a specific set of modifications to the superpotential
and the postulation of a few additional superfields.  In addition to
the matter fields of the MSSM, at least two massive vector-like
pairs of chiral superfields \(\Phi\) and \(\overline{\Phi}\) are
required, as are three generations of right-handed neutrino
superfield \(N_{a}\) and an additional exotic superfield \(\chi\),
whose function will be to acquire a scalar VEV.  An additional
symmetry, whose breaking will be responsible for late neutrino
masses,
is also imposed, and charges under it assigned so that the most
general leptonic-sector superpotential is
\begin{equation}
  \mathcal{W}\ni\lambda_{i\alpha}N_{\alpha}\Phi_{i}H_{u}+
  h_{i\alpha}L_{\alpha}\overline{\Phi}_{i}\chi+
  M_{\Phi_{i}}\Phi_{i}\overline{\Phi}_{i}+\mu H_{u}H_{d}.
  \label{eq:DiracLepSuperpotential}
\end{equation}
Here, \(\lambda\) and \(h\) are (generally complex) coupling
matrices and \(M_{\Phi_{i}}\) are the masses of the heavy
vector-like pairs, which, in order for successful baryogenesis to
occur, are generically required to be $10^{10}$~GeV or larger.\footnote{In general
$M_{\Phi}$, \(\lambda\), and \(h\) are all complex matrices, but we
can always choose to work in a basis where $M_{\Phi}$ is diagonal.}
Once the heavy superfields $\Phi_{i}$ and $\overline{\Phi}_{i}$ are
integrated out, the resulting effective superpotential
\begin{equation}
  \mathcal{W}_{\mathit{eff}}\ni
  \frac{\lambda_{i\alpha} h_{i\beta}^{\ast}}{M_{\Phi_{i}}}\chi L_{\beta} H_{u}
  N_{\alpha}
  +\mu H_{u} H_{d}
  \label{eq:EffDiracLepSuperpotential}
\end{equation}
will yield a small but nonzero Dirac mass term for the neutrinos,
provided that some mechanism is invoked to give the scalar component
of \(\chi\) a VEV.  Since \(\lambda\) and \(h\) are complex, the
decays of both the scalar and fermionic components of \(\Phi_{1}\)
and \(\overline{\Phi}_{1}\), the lightest of the additional heavy
superfields, will in general be CP-violating\footnote{It can be
shown that there is at least one nontrivial CP-violating phase in
\(\lambda\) and \(h\).} and will result in two equal and opposite
stores of lepton number, \(L_{\mathit{agg}}\) and \(L_{R}\).  The
first of these is an aggregate lepton number stored in left-handed
leptons, sleptons, and other fields in equilibrium with them, and is
transformed into baryon number by sphaleron interactions.  The
second is stored only in right-handed neutrinos,\footnote{We assume
  that right-handed sneutrinos couple strongly enough to
  other fields in the theory (e.g.\ with left-handed sneutrinos in
  the event that \(\chi\) has a large F-term VEV) to equilibrate with
  them and thus contribute to $L_{agg}$. Otherwise they would
  contribute to $L_R$.}
which, being singlets under \(SU(2)\times U(1)_{Y}\), do not
experience sphaleron effects and only couple to the other light
matter fields through the effective neutrino Yukawa interaction
given by~(\ref{eq:EffDiracLepSuperpotential}) with
\(\chi\rightarrow\langle\chi\rangle\). This interaction is
suppressed by \(\langle\chi\rangle/M_{\Phi_{1}}\) and therefore the
time scale for the equilibration of left-handed and right-handed
stores of lepton number 
can be quite late.  If the effective neutrino Yukawas are
sufficiently small,\footnote{Recall that in this scenario small
(Dirac) neutrino masses are
  obtained through small (effective) neutrino Yukawas.} the
equilibration time scale will be much longer than the time scale at
which sphaleron processes effectively shut off~\cite{Dick:1999je}.
When this is the case, a net baryon number for the universe will
already have frozen in 
and will persist unaltered until present time.

\indent In addition to providing a mechanism for baryogenesis, Dirac
leptogenesis holds some interesting implications for neutrino
physics.  Here, the squared neutrino masses are given by
\begin{equation}
  |m_{\nu}|^{2}_{\alpha\beta}= \Big(v \langle\chi\rangle \sin\beta\Big)^2\
  \sum_{i,j=1}^{2\ (or\ 3)} \sum_{\gamma=1}^3\lambda^{\ast}_{i\gamma}\lambda_{j\gamma}
  h^{\ast}_{i\alpha } h_{j\beta} \frac{1}{M^\ast_{\Phi_i}M_{\Phi_j}}.
  \label{eq:NeutMassSpelledOut}
\end{equation}
In order to reproduce the observed mass
splittings~\cite{Hagedorn:2005kz}
\begin{equation}
  \begin{array}{cc}
    \Delta m_{21}^{2}=\left(7.9^{+0.6}_{-0.6}\right)\times 10^{-5}
    \mathrm{eV}^{2},\ \ &
    |\Delta m_{31}^{2}|=\left(2.2^{+0.7}_{-0.5}\right)\times 10^{-3}
    \mathrm{eV}^{2},
  \end{array}
\end{equation}
where \(\Delta{m}_{ij}^{2}\equiv m_{i}^{2}-m_{j}^{2}\), and the
angles in the \(U_{\mathit{MNS}}\) matrix, it is necessary to impose
a few conditions on the coupling matrices \(\lambda\) and \(h\),
which determine the matrix structure
in~(\ref{eq:NeutMassSpelledOut}). There are several ways of doing
this, but we will focus on one particular, theoretically-motivated
scenario called constrained hierarchical Dirac leptogenesis
(CHDL)~\cite{Thomas:2005rs}, which produces a normal hierarchy among
neutrino masses.

\indent In CHDL, the appropriate matrix structure for \(\lambda\),
\(h\), and the (diagonal) mass matrix \(M_{\Phi}\) is obtained by
requiring that \(\lambda\) and \(h\) both be antisymmetric with
${\cal O}(1)$ entries and by relating \(M_{\Phi}\) to the fermion
Yukawas through a flavor symmetry. This structure ensures that the
neutrino mass matrix one obtains after integrating out the heavy
fields \(\Phi_{i}\) and $\overline{\Phi}_i$ corresponds to a normal
neutrino hierarchy, provided that there is a hierarchy among the
mass eigenstates \(M_{\Phi_{1}}\) and \(M_{\Phi_{2}}\).
Of course this coupling structure
also impacts baryogenesis, but it has been shown that in CHDL, the
constraints from baryogenesis and neutrino physics can be satisfied
simultaneously~\cite{Thomas:2005rs}.


\indent Baryogenesis and a small neutrino Dirac mass are not the
only consequences of equation~(\ref{eq:DiracLepSuperpotential}),
however.
Assuming that at some high scale $M$ all the soft supersymmetry
breaking terms are flavor diagonal and universal, the new terms of
equation~(\ref{eq:DiracLepSuperpotential}) will induce potentially
off-diagonal contributions to the slepton mass matrices
\(m^{2}_{LL}\), \(m^{2}_{RR}\) and \(m^{2}_{LR}\). As in Majorana
leptogenesis some of these contributions come from running the
scalar soft masses from some high scale \(M\) (the Planck scale, the
GUT scale, etc.) down to \(M_{\Phi_{1}}\). Assuming the Universality
condition at the high scale and that all soft A-terms are equal to
the relevant yukawa coupling multiplied by the universal soft
supersymmetric mass $m_s$, one can simply estimate the flavor
violating corrections to the mass matrix by integrating the RGE
equations iteratively~\cite{Hisano:1995cp,Petcov:2005jh}, and
generally obtain off-diagonal contributions to the slepton masses $\delta
m_{LL}^{2}$ and $\delta m_{LL}^{2}$:
\begin{eqnarray}
  \delta m_{LL}^{2}&\approx &
  -\frac{1}{2\pi^2}\ln\left(\frac{M}{M_{\Phi_{1}}}\right)h_{i\alpha}^{\ast}h_{i\beta}m_{s}^{2}\label{eq:MLL}\\
  \delta m_{RR}^{2}&\approx &
  -\frac{1}{2\pi^2}\ln\left(\frac{M}{M_{\Phi_{1}}}\right)\lambda_{i\alpha}^{\ast}\lambda_{i\beta}m_{s}^{2}\label{eq:MRR}.
\end{eqnarray}
If the F-term
of \(\chi\) acquires a VEV \(\langle F_{\chi}\rangle\) (a corollary
in most mechanisms via which its scalar component obtains a VEV),
the effective theory superpotential yields yet another potentially off
diagonal scalar mass
term
\begin{equation}
  \delta m_{LR}^{2}=h^{\dagger}_{i\alpha}\lambda_{i\beta}\frac{\langle
  F_{\chi}\rangle}{M_{\Phi_{1}}}v\sin\beta\label{eq:MLR}
\end{equation}
after electroweak symmetry breaking, which mixes left-handed and
right-handed sneutrinos.

\indent In CHDL, where we have a specific flavor structure for the
matrices \(\lambda\), \(h\) and \(M_{\Phi}\), one has a specific
prediction for flavor mixing among sleptons once the electroweak and
hidden symmetries are broken. In order to examine the effect of
these mixings, the full mass matrices for both the charged sleptons
and sneutrinos must be taken into account. For simplicity, we will
continue to assume that the leading soft breaking sector is flavor
diagonal and universal with a common scalar mass $m_s$. The
resulting additional contributions to the slepton mass squared
matrices, given by equations~(\ref{eq:MLL}),~(\ref{eq:MRR}),
and~(\ref{eq:MLR}),
can thus be expressed in terms of the \(3\times3\) submatrices
\(\delta m_{LL}^{2}\), \(\delta m_{RR}^{2}\), and \(\delta
m_{LR}^{2}\) as
\begin{equation}
  \delta m_{\tilde{\ell}^{\pm}}^{2}=
  {\small
  \left(\begin{array}{c:c}
  \delta m^2_{LL} & ~~0~~ \\[.25cm] \hdashline \\[-.25cm] ~~0~~ & ~~0~~
  \end{array}\right)}~~~~~~
  \delta m_{\tilde{\nu}}^{2}={\small
  \left(\begin{array}{c:c}
  \delta m^2_{LL} & \delta m^2_{LR} \\[.25cm] \hdashline \\[-.25cm] (\delta m^{2}_{LR})^\dagger & \delta m^2_{LL}
  \end{array}\right)}. \label{eq:MassMatricesDiracLep}
\end{equation}
The only contribution to the charged slepton mass squared matrix
comes from \(\delta m_{LL}^{2}\), while the sneutrino mass squared
matrix receives not only additional flavor mixings among left-handed
and among right-handed sneutrinos, but also an effective A-term from
\(\delta m_{LR}^{2}\) which intermixes left-handed and right-handed
sneutrinos.

\indent Once the matrices \(\lambda\) and \(h\) have been fixed, up
to an overall scaling parameter\footnote{In CHDL \(\lambda\) and
\(h\) are
  antisymmetric, with $\mathcal{O}(1)$ entries up to an overall scaling
  parameter f
 \begin{eqnarray}
   \lambda \approx f
  \left(\begin{array}{ccc} 0 & 1 & 1\\-1 & 0 & a_{3} \\ -1 & -a_{3} & 0\end{array}\right) &
   \hspace{.5cm} &
   h\approx f
   \left(\begin{array}{ccc} 0 & 1 & 1\\ -1 & 0 & b_{3} \\ -1 & -b_{3} & 0\end{array}\right),
   \label{eq:YukawaParametrization}
 \end{eqnarray}
where the values of $a_3$ and $b_3$ depend on the hierarchy between
\(M_{\Phi_{1}}\) and \(M_{\Phi_{2}}\): when
\(M_{\Phi_{2}}/M_{\Phi_{1}}=m_{\mu}/m_{e}\), $b_3$ can vary between
1.4 and 2.9, while $a_3$ can vary between 35 and 90; when
\(M_{\Phi_{2}}/M_{\Phi_{1}}=10\), $b_3$ can vary between 1.4 and
2.9, while $a_3$ can vary between 1.5 and 4.5 \cite{Thomas:2005rs}).}
$f$, to give the correct leptonic mixing matrix $U_{MNS}$, the
remaining free parameters of the model are \(\langle\chi\rangle\),
\(M_{\Phi_{i}}\), and \(f\). After the mass scale of the neutrinos
is fixed, these parameters are not independent anymore, and are
related by the constraint embodied in
equation~(\ref{eq:NeutMassSpelledOut}).
In CHDL, where
the lightest neutrino is very light relative to the other two, the
masses of the two heavier neutrinos are
\begin{eqnarray}
  m_{\nu_{2}}^{2}&\approx&\Delta m_{21}^{2}=
  \left(7.9^{+0.6}_{-0.6}\right)\times 10^{-5} \mathrm{eV}^{2} \\
  m_{\nu_{3}}^{2}&\approx&\Delta m_{31}^{2}=
  \left(2.2^{+0.7}_{-0.5}\right)\times 10^{-3} \mathrm{eV}^{2}.
\end{eqnarray}
Plugging in these values and using \(a_{3}=4.5\) and \(b_{3}=2.2\)
(the values most advantageous for baryogenesis with a chosen
hierarchy $M_{\Phi_2}/M_{\Phi_1}=10$) in the CHDL
parametrization~(\ref{eq:YukawaParametrization}) of \(\lambda\) and
\(h\), we find that the relation between the parameters
\(\langle\chi\rangle\), \(M_{\Phi_{i}}\), and \(f\), obtained from
the neutrino mass, is
\begin{equation}
 \frac{f^{2}\langle\chi\rangle}{M_{\Phi_{1}}}\sin\beta= 1.009\times10^{-13}.
\end{equation}
Using this relation, we can express the overall dependence of the
slepton mass terms in~(\ref{eq:MLL}), (\ref{eq:MRR}),
and~(\ref{eq:MLR}) on the relevant mass scales in the theory.
\begin{equation}
  \begin{array}{lcccc}
  \delta m_{LL}^{2}&\propto& f^2&\propto&
  \frac{M_{\Phi_{1}}}{\langle\chi\rangle}\\
  \delta m_{RR}^{2}&\propto& f^2& \propto&
  \frac{M_{\Phi_{1}}}{\langle\chi\rangle}\\
  \delta m_{LR}^{2}&\propto& \frac{f^2}{M_{\Phi_{1}}}&\propto&
  \frac {1}{\langle\chi\rangle}.
  \end{array}\label{eq:DeltaMassProps}
\end{equation}
Note that the proportionality constants for the bottom two equations
are not dimensionless: the ones associated with $\delta m_{LL}^{2}$
and $\delta m_{RR}^{2}$ each contain a factor of \(m_s^{2}\) and
have mass dimension $[m]^{2}$, while the one associated with $\delta
m_{LR}^{2}$ contains a factor of \(\langle F_{\chi}\rangle v\) and
has mass dimension $[m]^{3}$.

\section{Flavor Violation\label{sec:FlavorViolation}}

\indent

The most stringent constraints on flavor violation in the lepton
sector come from measurements of the branching ratios for
flavor-violating decays and conversions of heavy leptons, such as
\(\mu\rightarrow e\gamma\),  \(\tau\rightarrow \mu\gamma\),  $\mu\to
eee$ and  $\mu A\to eA$.  The current experimental limits on the
2-body decay processes are~\cite{Eidelman:2004wy}
\begin{eqnarray}
\mathit{BR}(\mu\rightarrow e\gamma)&<&1.2\times 10^{-11},\label{eq:CurrentExpConstraints}\\
\mathit{BR}(\tau\rightarrow \mu\gamma)&<&1.1\times 10^{-6}.
\end{eqnarray}
In the near future, the MEG experiment~\cite{MEGHome:2006mh}
is expected to
improve the current experimental bound on \(\mu\rightarrow e\gamma\)
by several orders of magnitude, to \(\mathcal{O}(10^{-13}-10^{-14})\)
or lower.  Other related projects, such as PRIME~\cite{Kuno:2005mm}
(sensitive to \(\mu A\rightarrow e A\) conversion), are expected to
go online over the next few years.  Projects have also been
proposed~\cite{Calibbi:2006nq} that would lower the bound from
\(\tau\rightarrow \mu\gamma\) to \(\mathcal{O}(10^{-9})\).

\begin{figure}[ht!]
\begin{center}
\includegraphics[width=10cm]{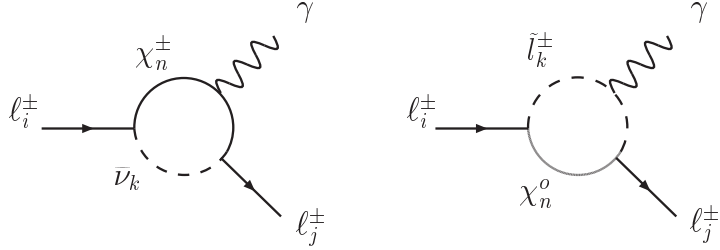}
\end{center} \caption{Feynman diagrams giving the two leading order
contributions to the flavor-changing process
\(\ell_{i}^{-}\rightarrow\ell_{j}^{-}\gamma\) due to sneutrino (left
diagram) and charged slepton (right diagram) mass
mixings.\label{fig:FeynMixingsFull}}
\end{figure}

\indent The effective interaction leading to lepton flavor
  violating decays of the form
\(\ell_{i}\rightarrow\ell_{j}\gamma\), where $\ell_{i}$ and
$\ell_{j}$ are charged leptons, can be written as
\begin{equation}
{\cal I}=i\hspace{.125em}e\hspace{.125em} m_{\ell_j}\hspace{.125em}
\bar{u}_i(q-p)\hspace{.125em}
\sigma_{\alpha\beta}\hspace{.125em}q^\beta\left(A^L P_L+A^R
P_R\right)u_j(p)\hspace{.125em} \epsilon^*(q),
\end{equation}
where $q$ and $p$ are the momenta of the photon and the outgoing
lepton $\ell_j$ respectively, and $m_{\ell_j}$ is the outgoing
lepton mass.  The resulting decay rate is
\begin{eqnarray}
\Gamma(l_j^- \rightarrow l_i^-~\gamma) = \frac{e^2}{16 \pi}
m_{l_j}^5 (|A^L|^2+|A^R|^2).
\label{eq:ViolationRate}
\end{eqnarray}
The leading contributions to the amplitudes $A^L$ and $A^R$ appear
at one loop level and are shown in figure~\ref{fig:FeynMixingsFull}.
They involve both a sneutrino (and chargino) mass eigenstate and
charged slepton (and neutralino) mass eigenstate running in the
loop. These amplitudes have been computed in~\cite{Hisano:1995cp}
for a general MSSM scenario\footnote{In our case we need to add
three right handed sneutrinos, but it is trivial to extend the
result to include six sneutrino mass eigenstates instead of three.}
but for completeness we include them in Appendix A.

\indent In order to proceed further, it will be necessary to make a
few assumptions concerning the supersymmetric model parameters.
Those relevant to a discussion of lepton-sector flavor violation
include the gaugino masses $M_1$ and $M_2$, the Higgs mass parameter
\(\mu\), the ratio of Higgs VEVs \(\tan\beta\), and the soft masses
for the sleptons.
In our analysis, we choose the values \(M_{1}=160\)~GeV,
\(M_{2}=220\)~GeV, \(\mu=260\)~GeV, and \(\tan\beta=3, 10\) and $30$. As for the
slepton soft masses
we will, as previously mentioned, assume a common scale $m_{s}$ and
examine what effect varying \(m_{s}\) has on
\(\mathit{BR}(\mu\rightarrow e\gamma)\) and
\(\mathit{BR}(\tau\rightarrow \mu\gamma)\).  We will assume that the
scale at which soft masses are universal is \(M=2\times10^16\)~GeV,
though the results are not particularly sensitive to this choice.
\begin{figure}[ht!]
\begin{center}
  \includegraphics[height=11cm,width=7.5cm,trim=0cm .65cm 0cm 0cm]{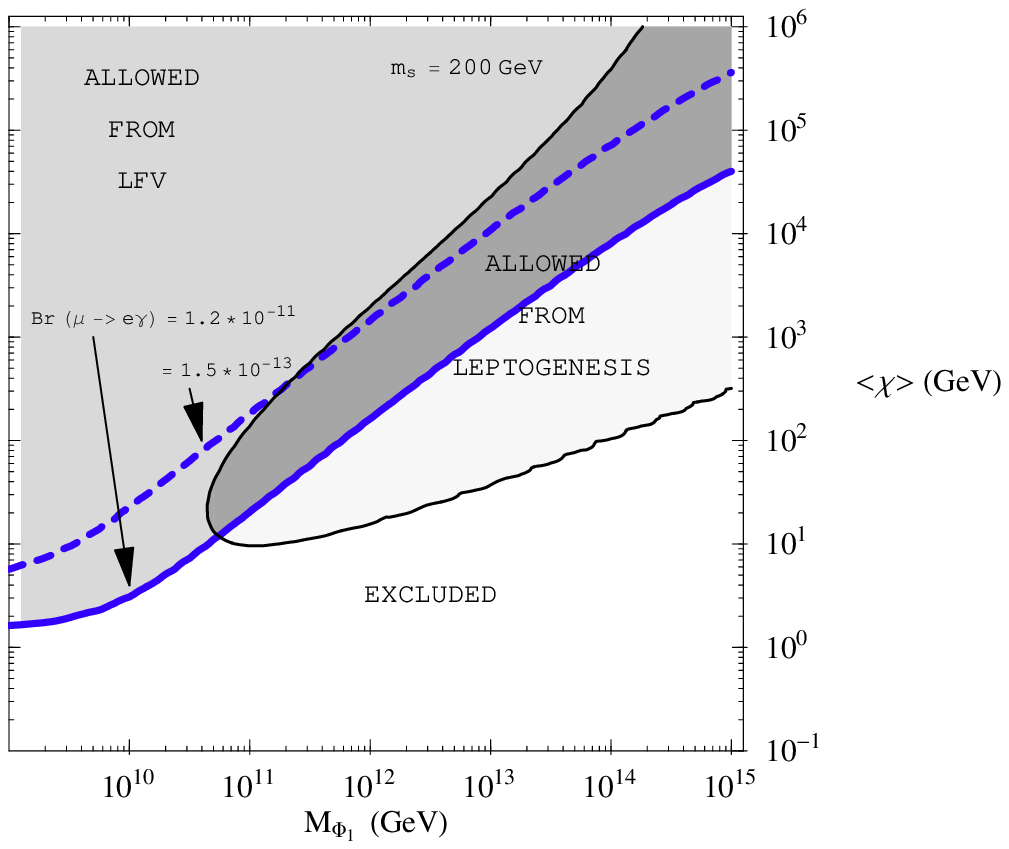}
\hspace{.5cm}
\includegraphics[height=10.5cm,width=7.3cm,trim=0cm 0cm 0cm 0cm]{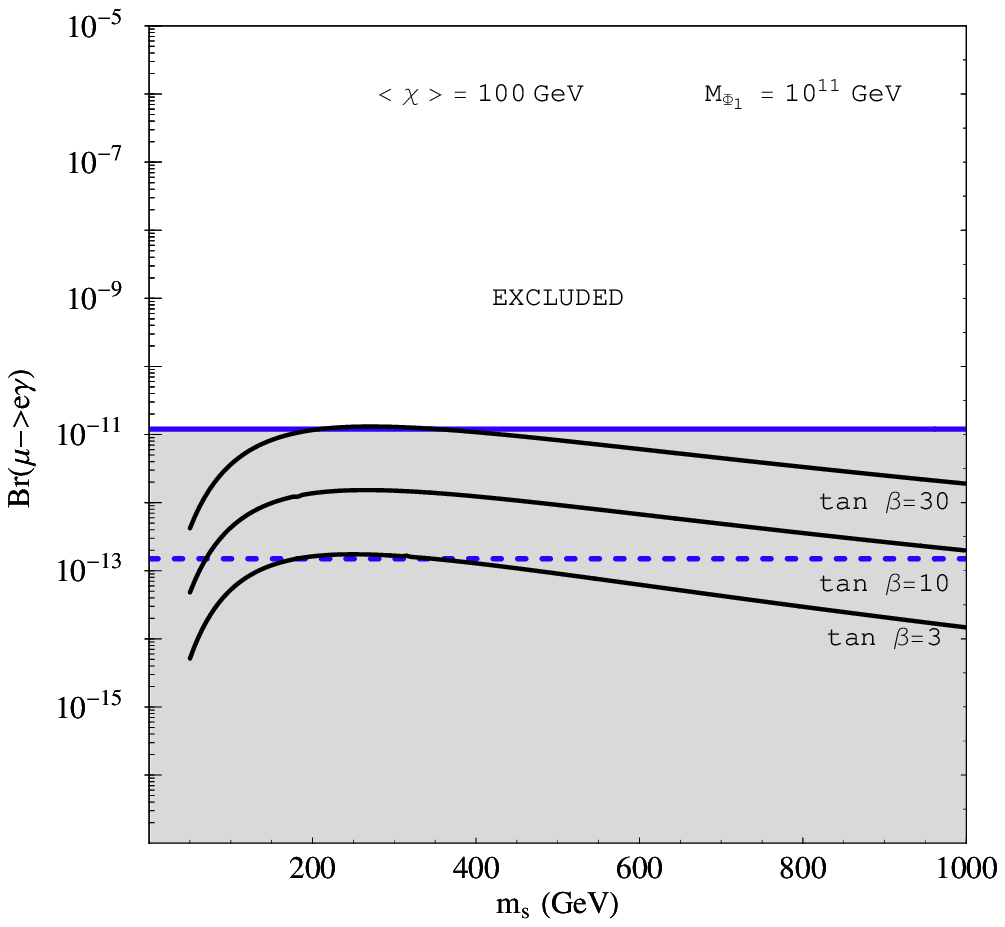}
\end{center}
\vspace{-1cm} \caption{Exclusion plots combining constraints from
both leptogenesis and flavor violation in the process
\(\mu\rightarrow e\gamma\).  The left-hand panel shows exclusion
contours in \(M_{\Phi_{1}}\)-\(\langle\chi\rangle\) space for a
universal scalar soft mass $m_{s}=200$~GeV, with $\tan\beta=10$; the right hand panel
shows the variation of the branching ratio
\(\mathit{BR}(\mu\rightarrow e\gamma)\) with respect to $m_{s}$ using
$\tan\beta=3, 10$ and $30$.
  In both plots, we have
taken \(M_{1}=160\)~GeV, \(M_{2}=220\)~GeV, and
\(\mu=260\)~GeV.  We have also assigned the \(\chi\) superfield an
F-term VEV $\sqrt{\langle
  F_\chi\rangle}=10^7$ GeV.  Such a large VEV results in large trilinear couplings
between Higgs fields and sneutrinos and therefore induces
potentially sizeable mixings between left-handed and right-handed
sneutrinos after electroweak symmetry breaking.  In each plot, the
thick solid contours represent the current experimental bound on the
branching fraction~(\ref{eq:CurrentExpConstraints}); the dashed
lines represent the expected future experimental bound of
$1.5\times10^{-13}$ from MEG.
The thin solid contour in the left-hand panel delimits the region
allowed by leptogenesis constraints.\label{fig:ExclusionPlot}}
\end{figure}

\indent The results of our calculation are displayed in
figure~\ref{fig:ExclusionPlot}.  In the left panel, we show
exclusion contours in \(M_{\Phi_{1}}\)-\(\langle\chi\rangle\) space
for $m_{s}=200$~GeV. The areas below and to the right of the lower
contour (the white region) are excluded by the experimental bounds
given in~(\ref{eq:CurrentExpConstraints}).  We also include contours
demarcating the region wherein baryogenesis can succeed, which have
been updated from~\cite{Thomas:2005rs} to include the
effects\footnote{We thank A.~Strumia for pointing out to
  us the potential numerical importance of these thermalization processes involving
  gauge interactions\cite{Hambye:2005tk,Chun:2005ms}.} of processes second order in \(\Phi_{1}\) and
\(\overline{\Phi}_{1}\) (see Appendix~\ref{app:SecondOrderPhi}).
Contours have also been computed for \(\tau\rightarrow\mu\gamma\),
but the constraints they imply for the theory are far weaker than
those from \(\mu\rightarrow e\gamma\).
In the right panel, we show how varying the universal scalar mass
affects the branching ratio for \(\mu\rightarrow e\gamma\), which
reaches a maximum when \(m_{s}\) is around the weak scale.  This is
to be expected: when \(m_{s}\) is much larger than the weak scale
both the slepton mass-squared eigenvalues and the flavor-violating
terms scale like $m_{s}^{2}$ and the sneutrino and charged slepton
mixing matrices asymptote to a constant value, while the branching
ratio is still suppressed by the masses running in the loop; as
$m_{s}$ decreases below the weak scale, \(\delta m_{LL}^{2}\) and
\(\delta m_{LL}^{2}\) go to zero and the slepton masses are
dominated by flavor diagonal electroweak contributions. We also
observe that, as in the SUSY see-saw case~\cite{Hisano:1995cp,Petcov:2005jh}, the flavor violation rate
is quite sensitive to \(\tan\beta\).

\begin{figure}[ht!]
\vspace{1cm}
\begin{center}
\includegraphics[width=14cm]{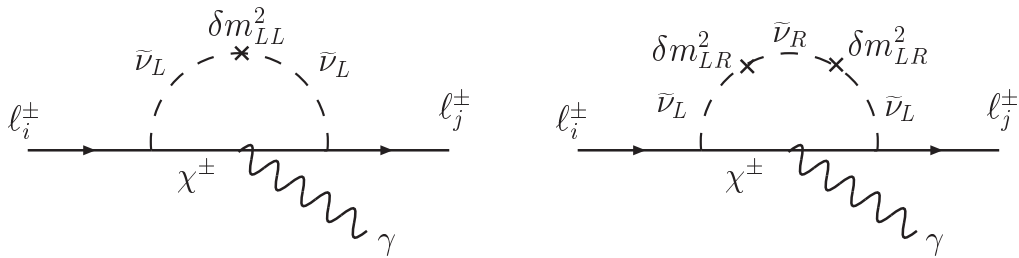}
\end{center} \caption{Feynman diagrams for the leading-order process involving \(\delta m^{2}_{LL}\)
(left diagram), and for the leading process involving \(\delta
  m^{2}_{LR}\) (right diagram),
with sneutrinos running in the loop in the mass-insertion
approximation.  Note that any process involving \(\delta
m^{2}_{LR}\) necessarily involves two mass insertions, and any one
involving \(\delta m^{2}_{RR}\) (given by the diagram on the right
  with an additional \(\delta m^{2}_{RR}\) insertion)
necessarily involves three.
\label{fig:FeynMixingsMassInsert}}
\end{figure}

\indent In interpreting the results in
figure~\ref{fig:ExclusionPlot}, it is useful to note that in the
regions of the plot near the exclusion contours (where the branching
ratio for \(\mu\rightarrow e\gamma\) is quite low), flavor-violating
effects will be small. It is therefore valid to use the
mass-insertion approximation there and treat \(\delta m^{2}_{LL}\)
(left), \(\delta m^{2}_{LR}\), and \(\delta m^{2}_{RR}\) as small
corrections to the slepton propagators.  Since the sneutrino
propagator can receive mass insertions from all three, we will focus
our analysis on the partial amplitude from the sneutrino-mediated
process (the left diagram in figure~\ref{fig:FeynMixingsFull}).  In
figure~\ref{fig:FeynMixingsMassInsert}, we list the leading
contributions to this partial amplitude involving each of \(\delta
m^{2}_{LL}\) (left diagram), \(\delta m^{2}_{LR}\), and \(\delta
m^{2}_{RR}\).  Since there is no coupling between leptons and
right-handed sneutrinos, corrections from \(\delta m^{2}_{LR}\) and
\(\delta m^{2}_{RR}\) only appear at second and third order in the
mass insertion expansion, respectively.  Therefore, if there is no
substantial hierarchy among these three sets of mixing terms, the
primary source of flavor-violation comes from mixings between
left-handed sleptons.  In the mass insertion approximation, the
branching ratio for such processes can naively be estimated as
\begin{equation}
  \mathit{BR}(\mu\rightarrow e\gamma)\propto
  \frac{\alpha^{3}}{G_{F}^{2}} \frac{(\delta
  m_{LL}^{2})^{2}}{m_{s}^{8}}
  \label{eq:BRwithLefts}
\end{equation}
and therefore contours of branching ratio in the
$M_{\Phi_1}-\langle\chi\rangle$ plane correspond to contours of
$\delta m_{LL}^{2}$.  According to
equation~(\ref{eq:DeltaMassProps}), $\delta
m_{LL}^{2}=c_{1}M_{\Phi_1}/\langle\chi\rangle$, where $c_{1}$ is a
dimensionless proportionality constant with dimension \([m]^2\), so
the exclusion contour associated with left-left mixing takes the
form
\begin{equation}
\ln M_{\Phi}=\ln\langle\chi\rangle+C_{LL},
\label{eq:ContourEquationLL}
\end{equation}
where \(C_{LL}=-\ln(\delta m_{LL}^{2}/c_{1})\).  The oblique, upper
exclusion contour in figure~\ref{fig:ExclusionPlot}, which embodies
the constraint from left-left mixing, is associated with this linear
equation.

\indent While some hierarchy among the mass insertion terms is
necessary for \(\delta m^{2}_{LR}\) and \(\delta m^{2}_{RR}\) to be
relevant, there is no a priori reason why such a hierarchy should
not exist.  The \(\delta m^{2}_{LR}\) contribution~(\ref{eq:MLR}) is
proportional to \(\langle F_{\chi}\rangle\), which is essentially a
free parameter.  As mentioned above, \(\langle F_{\chi}\rangle\) is
not even relevant to Dirac leptogenesis per se, but appears as a
common side-effect of mechanisms for assigning the \(\chi\)
superfield a scalar VEV. Still, in many such
mechanisms~\cite{Borzumati:2000mc}, the scale \(\sqrt{\langle
F\rangle}\) can potentially be quite large (\(10^{6}~\mathrm{GeV}\)
or higher), and if this is the case, contribution from \(\delta
m^{2}_{LR}\) could potentially be as important as those from
\(\delta m^{2}_{LL}\).  Let us assume for a moment that this is the
case and examine the constraints related to \(\delta m^{2}_{LR}\)
and \(\delta m^{2}_{LR}\) together. In regions of
figure~\ref{fig:ExclusionPlot} where \(\langle\chi\rangle\) is
small, we now have
\begin{equation}
BR(\mu\rightarrow e\gamma)\propto
\frac{\alpha^{3}}{G_{F}^{2}m_{S}^{8}} \frac{(\delta
m^{2}_{LR})^{4}}{m_{\tilde{\nu}_{R}}^{4}}=\mathrm{constant}
\label{eq:BRwithRights}
\end{equation}
along any exclusion contour.  Equation~(\ref{eq:DeltaMassProps})
tells us that,
\begin{equation}
  \delta m_{LR}^2=c_{2}\frac{1}{\chi},
\end{equation}
where $c_{2}$ has mass dimension $[m]^{3}$.  Therefore, when $\delta
m^{2}_{RR}\ll m_{\tilde{\nu}_{R}}^{2}$,
the associated contour is given by
\begin{equation}
\ln\langle\chi\rangle=C_{LR} \label{eq:ContourEquationLR},
\end{equation}
where \(C_{LR}=\ln(c_{2}/\delta m_{LR}^{2})\).
This equation explains the behavior of the contour in the left panel
of figure~\ref{fig:ExclusionPlot} when \(\langle\chi\rangle\) is
small and the oblique bound from left-left mixing abruptly ells into
a horizontal line---the bound from left-right mixing.
In general, the effect of increasing \(\langle F_{\chi}\rangle\) is
to push this latter bound upward, and for \(\sqrt{\langle
F_{\chi}\rangle}\gtrsim 10^{9}\), the entirety of parameter space is
excluded.  The effect of right-right mixing is higher order still
and only becomes relevant in regions of parameter space where
\(M_{\Phi_{1}}\) is large and \(\langle\chi\rangle\) is
small---regions already excluded by left-left mixing.

For the numerical analysis shown in Figure~\ref{fig:ExclusionPlot},
we have taken $\sqrt{\langle
  F_{\chi}\rangle}=10^7$ GeV which in some parts of the
$\langle\chi\rangle-M_{\Phi_1}$ plane, induces substantial mixing
between left-handed and right-handed sneutrinos.  Sizeable mixing of
this sort opens the intriguing possibility that the lightest
sneutrino could be a legitimate dark matter
candidate~\cite{Hall:1997ah,Arkani-Hamed:2000bq,Hooper:2004dc,Asaka:2005cn} in
Dirac leptogenesis,
a possibility which would be interesting to investigate in the
future.

\indent The primary message of figure~\ref{fig:ExclusionPlot} is
that the combined constraints from leptogenesis and flavor violation
do not exclude Dirac leptogenesis in theories with low-scale
sfermion masses.  In general, the latter set of constraints tend to
rule out theories with exceptionally high masses for the decaying
particles \(\Phi_{1}\) and \(\overline{\Phi}_{1}\) and exceptionally
low values for \(\langle\chi\rangle\).\footnote{Recall that the
parameters $\langle \chi \rangle$, $M_{\Phi_1}$ and $f$ enter in the
effective neutrino Yukawa coupling given by
\ref{eq:EffDiracLepSuperpotential}, so when one of them is
increased, we have to appropriately tune the other two to maintain
the neutrino Yukawas at a fixed value.  Such an adjustment, in turn,
changes the different contributions from these parameters to
off-diagonal entries in the slepton mass matrices.}  Given current
bounds on lepton flavor violation, there is still a sizeable region
of parameter space within which the scenario succeeds, even when
\(m_{s}\) is as low as \(100\)~GeV.  It is of interest, however,
that MEG and the next generation of lepton flavor violation
experiments will be able to probe the vast majority of this region,
and the data from these experiments will be crucial in determining
the viability of Dirac leptogenesis with weak-scale superpartners.

\indent
As was mentioned in
section~\ref{sec:DiracLep}, some new symmetry must be posited in
order to obtain late neutrino mass generation. 
In Dirac leptogenesis, neutrino masses are the result of the scalar
component of the \(\chi\) superfield acquiring a VEV which breaks
this new symmetry,
producing a Goldstone boson or pseudo-Goldstone boson, depending on
the nature of the symmetry.  Constraints on such bosons arise from
both BBN and cosmic microwave background (CMB)
considerations~\cite{Hall:2004yg} as well as from the detection of
abnormalities in the neutrino flux associated with supernova
events~\cite{Goldberg:2005yw}, and they can become problematic
(depending on the mass of the Goldstone boson) when the
symmetry-breaking VEV is less than around 1~GeV.  The value of
\(\langle\chi\rangle\) required by leptogenesis constraints is
around 10~GeV, and thus the cosmological complications associated
with breaking the additional symmetry necessary for Dirac
leptogenesis do not pose any problem for the theory.

\section{Conclusion\label{sec:Conclusion}}

\indent

It has already been shown~\cite{Thomas:2005rs} that Dirac
leptogenesis stands as a phenomenologically viable alternative to
the standard leptogenesis picture.  In theories with comparatively
light scalars, however, flavor violation becomes a serious concern
commonly solved by the Universality assumption. The Dirac
leptogenesis superpotential~(\ref{eq:DiracLepSuperpotential}) gives
rise to new mass terms for sneutrinos and charged sleptons which
induce mixings between flavor eigenstates after the breaking of both
the electroweak symmetry and the new symmetry responsible for late
neutrino mass generation.
 Experimental limits on flavor violation
in heavy lepton decays such as \(\mu\rightarrow e\gamma\)
significantly constrain any theory which permits slepton flavor
mixing, and in Dirac leptogenesis these constraints translate
into bounds on the theory parameters \(M_{\Phi_{1}}\) (the mass of
the heavy decaying particle) and \(\langle\chi\rangle\) (the VEV of
the exotic scalar field). Baryogenesis requirements also place
significant constraints on both of these parameters, and thus the
question as to whether leptogenesis can be made to work at all when
scalar masses are light is a highly nontrivial one.

\indent In this work, we have shown that even when the masses of
supersymmetric scalars are small, substantial regions of parameter
space exist for which Dirac leptogenesis succeeds in producing a
realistic baryon asymmetry for the universe while respecting current
bounds on flavor violation.  This is true even when the masses of
supersymmetric particles are as low as \(\sim 100\)~GeV.
Interestingly enough, for such light fields, which presumably can be
discovered at the LHC, it is generically predicted that experiments
such as MEG should have enough sensitivity to observe flavor
changing effects if Dirac leptogenesis is in fact responsible for
the baryon asymmetry of the universe.

\section{Acknowledgments}

\indent

We would like to thank James Wells for his useful comments and
discussions and for carefully reading the manuscript. We also thank
David Morrissey  and Kazuhiro Tobe for useful
comments or related discussions.  B.T. and M.T. are supported by
D.O.E. and the Michigan Center for Theoretical Physics (MCTP).


\appendix
\section{Effective Couplings\label{app:EffectiveCouplings}}

\indent

For completeness, we list here the results used in
our analysis for lepton flavor violating processes.
The amplitudes $A^{L}$ and $A^{R}$ in
equation~(\ref{eq:ViolationRate}) were computed in
\cite{Hisano:1995cp} and, with a trivial extension to include three right-handed
sneutrinos, are given by
\begin{equation}
  A^{L}=A^{(c),L}+A^{(n),L}\hspace{0.75cm}\mathrm{and}\hspace{0.75cm}A^{R}=A^{(c),R}+A^{(n),R},
\end{equation}
where the individual amplitudes $A^{(c),L}$, $A^{(n),L}$,
$A^{(c),R}$, and $A^{(n),R}$ are
\begin{eqnarray}
A^{(n)L}&=&\frac{1}{32
\pi^2}\sum_{A=1}^{4}\sum_{X=1}^{6}\frac{1}{m^2_{\tilde{\ell}_X}}
\left[
 N_{iAX}^{L} N_{jAX}^{L*}
\frac{1}{6 (1-x_{AX})^4} \right.
\nonumber \\
&&\times (1-6x_{AX}+3x_{AX}^2+2x_{AX}^3-6x_{AX}^2\ln x_{AX})
\nonumber \\
&&\left. +N_{iAX}^{L} N_{jAX}^{R*}
\frac{M_{\tilde{\chi}_A^0}}{m_{l_j}} \frac{1}{(1-x_{AX})^3}
(1-x_{AX}^2+2x_{AX} \ln x_{AX}) \right], \label{eq:ANL}
\\
A^{(c)L}&=&-\frac{1}{32 \pi^2}
\sum_{A=1}^{2}\sum_{X=1}^{6}\frac{1}{m^2_{\tilde{\nu}_X}} \left[
 C_{iAX}^{L} C_{jAX}^{L*}
\frac{1}{6 (1-x_{AX})^4} \right.
\nonumber \\
&&\times (2+3x_{AX}-6x_{AX}^2+x_{AX}^3+6x_{AX} \ln x_{AX})
\nonumber \\
&&\left. +C_{iAX}^{L} C_{jAX}^{R*}
\frac{M_{\tilde{\chi}_A^-}}{m_{l_j}} \frac{1}{(1-x_{AX})^3}
(-3+4x_{AX}-x_{AX}^2-2 \ln x_{AX}) \right],\label{eq:ACL}\\
A^{(n,c)R}&=&A^{(n,c)L}|_{L \leftrightarrow R}\label{eq:ACRANR}.
\end{eqnarray}
Here, the indices \(A\) and \(X\) respectively label the gaugino
(chargino or neutralino) and slepton (sneutrino or charged slepton)
mass eigenstates, $x_{AX}\equiv m^{2}_{\chi_{A}}/m^{2}_{\phi_{X}}$,
and $C_{iAX}^{L,R}$ ($N_{iAX}^{L,R}$) denote the effective couplings
of charged lepton \(i\) to chargino (neutralino) \(A\) and
sneutrino (charged slepton) \(X\). The flavor mixing terms
in~(\ref{eq:MassLagrangian}) enter into the overall
rate~(\ref{eq:ViolationRate}) through $C_{iAX}^{L,R}$ and
$N_{iAX}^{L,R}$, which contain elements of the matrices $U_{\nu}$
and \(U_{\ell}\) that diagonalize the mass-squared matrices for
sneutrinos and charged sleptons, respectively.  The slepton masses
also enter into the partial
amplitudes~(\ref{eq:ANL}-\ref{eq:ACRANR}).

The effective couplings $N^{L,R}_{iAX}$ and $C^{L,R}_{iAX}$ are
\begin{eqnarray}
  N^{R}_{iAX}&=& -\frac{g_2}{\sqrt{2}} \left(
       [-(U_N)_{A2} -(U_N)_{A1} \tan \theta_W] U^{\ell}_{X,i}
        + \frac{m_{l_i}}{m_W\cos\beta} (U_N)_{A3} U^{\ell}_{X,i+3}
        \right),
\nonumber \\
  N^{L}_{iAX} &=& -\frac{g_2}{\sqrt{2}} \left(
           \frac{m_{l_i}}{m_W\cos\beta} (U_N)_{A3}
           U^{\ell}_{x,i}
           +2 (U_N)_{A1} \tan \theta_W U^{\ell} _{X,i+3} \right),
\nonumber\\
 C^{R}_{iAX}& =& -g_2(O_R)_{A1} U^{\nu}_{X,i},~~~\mathrm{and}
\nonumber \\
 C^{L}_{iAX}& = & g_2\frac{m_{l_i}}{\sqrt{2}m_W\cos\beta}(O_L)_{A2}
                    U^{\nu}_{X,i}
\end{eqnarray}
in terms of the chargino mixing matrices $(O_R)_{A,i}$ and
$(O_L)_{A,i}$ the neutralino mixing matrix $U^{N}_{X,i}$, and the
sneutrino and charged slepton mixing matrices $U^{\nu}_{X,i}$ and
$U^{\ell}_{X,i}$.  The chargino mixings matrices are defined by the
relation
\begin{equation}
 M_{c}^{\mathit{diag}}=(O_R)M_{c}(O_L)^{T},
\end{equation}
where
\begin{equation}
  M_{c}=\left(\begin{array}{cc} 0 & X \\ X^{T} & 0
  \end{array}\right),~~~\mathrm{where}~~~ X=\left(\begin{array}{cc} M_{2} & \sqrt{2}M_W\cos\beta \\
  \sqrt{2}M_{W}\sin\beta & \mu \end{array}\right)
\end{equation}
and $M_{c}^{\mathit{diag}}$ is diagonal.  The sneutrino mixing
matrix $U^{\nu}_{X,i}$ and the charged slepton mixing matrix
$U^{\ell}_{X,i}$ are defined by the relations
\begin{equation}
  (m_{\tilde{\ell}^{\pm}}^{2})^{\mathit{diag}}=U^{\ell}m_{\tilde{\ell}^{\pm}}^{2}U_{\ell}^{\dagger}~~~,~~~
  (m_{\tilde{\nu}^{\pm}}^{2})^{\mathit{diag}}=U^{\ell}m_{\tilde{\nu}}^{2}U_{\ell}^{\dagger},
\end{equation}
where the matrices $m_{\tilde{\ell}^{\pm}}^{2}$ and
$m_{\tilde{\nu}}^{2}$ are given by the sum of the MSSM contribution
 and the respective Dirac leptogenesis
contributions in~(\ref{eq:MassMatricesDiracLep}).  The neutralino
mixing matrix $U_{N}$ is defined by the relation
\begin{equation}
  (m_{\tilde{N}})^{\mathit{diag}}=U_{N}m_{\tilde{N}}U_{N}^{\dagger},
\end{equation}
where
\begin{equation}
  m_{\tilde{N}}=\left(\begin{array}{cccc}
  M_1 & 0 & -M_Z\sin\theta_w\cos\beta & M_Z\sin\theta_w\sin\beta \\
  0 & M_2 & M_Z\cos\theta_w\cos\beta & -M_Z\cos\theta_w\cos\beta \\
  -M_Z\sin\theta_w\cos\beta & M_Z\cos\theta_w\cos\beta & 0 & -\mu \\
  M_Z\sin\theta_w\sin\beta & -M_Z\cos\theta_w\sin\beta & -\mu &
  0 \end{array}\right).
\end{equation}

\section{Boltzmann Equations Including Second Order Processes\label{app:SecondOrderPhi}}

\indent

In~\cite{Thomas:2005rs}, we derived the system of Boltzmann
equations for Dirac leptogenesis up to processes first order in the
heavy, decaying particles \(\Phi_{1}\) and \(\overline{\Phi}_{1}\)
and showed that the dynamics of these fields could be expressed
using only two equations: one for the lepton number
\(L_{\phi_{\Phi}}\) in the heavy field sector and one for the
abundance $Y^{c}_{\phi_{\Phi}}$ of the scalar component
\(\phi_{1}\) of the \(\Phi_{1}\) supermultiplet.  Here, we improve
upon our previous calculations by including terms second order in
\(\phi_{1}\).  These terms only appear in the equation for the
$Y^{c}_{\phi_{\Phi}}$ abundance, which becomes\footnote{Since the
dynamics of scalar and fermion fields in \(\Phi_{1}\) and
\(\overline{\Phi}_{1}\) are assumed to be the same, contributions of
the form \(\tilde{\phi}_{1}\phi_{1}\rightarrow ij\) can be
incorporated into $\gamma_{A}$.}
\begin{equation}
  \frac{dY^{c}_{\phi_{\Phi}}}{dz}=-\gamma_{D}
  \left(\frac{Y^{c}_{\phi_{\Phi}}}{Y^{eq}_{\phi_{\Phi}}}-1\right)+
  \frac{1}{2}\gamma_{L}^{\mathit{ID}}\frac{L_{\mathit{agg}}}{Y^{eq}_{\phi_{\Phi}}}+
  \frac{1}{2}\gamma_{R}^{\mathit{ID}}\frac{L_{\nu_{R}}}{Y^{eq}_{\phi_{\Phi}}}-
  \gamma_{A}\left(\frac{(Y^{c}_{\phi_{\Phi}})^{2}}{(Y^{eq}_{\phi_{\Phi}})^{2}}-1\right),
  \label{eq:dYcBoltz}
\end{equation}
where  $L_{\nu_{R}}$ is the lepton number stored in sneutrinos,
$L_{\mathit{agg}}$ is the aggregate lepton number stored in the
other, interacting fields that carry lepton number, and all
abundances are normalized with respect to the entropy density $s$.
The reaction densities \(\gamma_{D}\), \(\gamma_{L}^{\mathit{ID}}\),
and \(\gamma_{R}^{\mathit{ID}}\) are given by
\begin{eqnarray}
  \gamma_{D}&=&\left(\frac{K_{1}(z)}{K_{2}(z)}\right)
    s Y^{c}_{\phi_{\Phi}} \Gamma_{D}\\
  \gamma_{L}^{\mathit{ID}}&=&\frac{1}{7}\frac{n^{eq}_{\phi_{\Phi}}}{n_{\gamma}}
    \left(\frac{K_{1}(z)}{K_{2}(z)}\right) s Y^{c}_{\phi_{\Phi}}
    \Gamma_{L}\\
  \gamma_{R}^{\mathit{ID}}&=&\frac{n^{eq}_{\phi_{\Phi}}}{n_{\gamma}}
    \left(\frac{K_{1}(z)}{K_{2}(z)}\right) s Y^{c}_{\phi_{\Phi}}
    \Gamma{R},
\end{eqnarray}
in terms of the Bessel functions $K_{1}(z)$ and $K_{2}(z)$, the
photon number density \(n_{\gamma}\), and the rates $\Gamma_{D}$,
$\Gamma_{L}$, and $\Gamma_{R}$, which respectively represent the
total decal width of the scalar component field \(\phi_{1}\) and the
individual decay widths for the processes
\(\phi\rightarrow\tilde{\ell}\chi\) and
\(\phi\rightarrow\nu^{c}_{R}\tilde{H}^{c}_{u}\).  The reaction
density \(\gamma_{A}\), which is associated with second order
processes of the form \(\phi_{1}\phi_{1}\rightarrow ij\) and
\(\tilde{\phi}_{1}\phi_{1}\rightarrow ij\), is given by
\begin{equation}
  \gamma_A=\frac{T}{64\pi^{4}}\int^{\infty}_{s_{\mathit{min}}}s^{1/2}
  K_{1}\left(\frac{\sqrt{s}}{T}\right)\hat{\sigma}(s),
\end{equation}
where $T$ is temperature, $s$ is the usual Mandelstam variable, and
\(\hat{\sigma}(s)\) is the total reduced cross section for
annihilations of \(\phi_{1}\)\(\phi_{1}\) and \(\phi_{1}\)\(\tilde{\phi}_{1}\)
into light fields.  This is defined by the formula
\begin{equation}
  \hat{\sigma}(s)=\frac{1}{8\pi s}\int^{t_{+}}_{t_{-}}
  \sum_{i}|\mathcal{M}_{i}(t)|^{2}dt,
\end{equation}
where both $t$ and $s$ denote the Mandelstam variables. The limits of
integration are given by $t_\pm=M_{\Phi_1}^2-{s}(1\mp r)/2$, with
$r$ defined below.

\begin{figure}[ht!]
\begin{center}
\includegraphics[width=7cm,height=9.5cm]{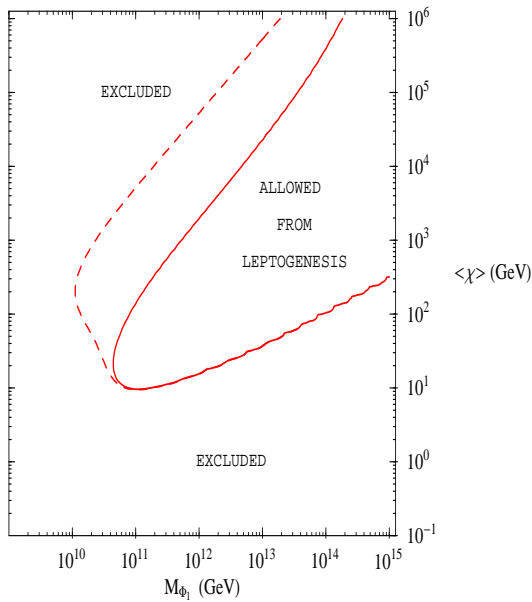}
\vspace{-1cm}
\end{center} \caption{This figure illustrates the effect of
second-order processes of the form \(\phi_{1}\phi_{1}\rightarrow
ij\) and \(\tilde{\phi}_{1}{\phi}_{1}\rightarrow ij\) on the
exclusion contours from leptogenesis. Contours are displayed for the
case without annihilation (dashed line) and with annihilation (solid
line). \label{fig:SecondOrderFX}} \vspace{.5cm}
\end{figure}

\indent In supersymmetric Dirac leptogenesis, the total reduced
cross-section \(\gamma_{A}\), including all relevant decay
processes, is
\begin{eqnarray}
  \hat{\sigma}_{\mathit{SUSY}}^{\mathit{tot}}&=&
  \frac{1}{16\pi}\left[
  6g_{Y}^{2}g_{2}^{2}\left(\left(-7+\frac{4}{x}\right)r
    +\left(\frac{8}{x^2}-\frac{4}{x} +
    9\right)\ln\left(\frac{1+r}{1-r}\right)\right)\right.\nonumber\\&&+
  g_{2}^{4}\left(\left(32+\frac{66}{x}\right)r+
    3\left(-\frac{16}{x^{2}}-\frac{16}{x}+9\right)
    \ln\left(\frac{1+r}{1-r}\right)\right)\nonumber\\&&+\left.
  g_{Y}^{4}\left(\left(19-\frac{36}{x}\right)r+
    \left(\frac{16}{x^2}-\frac{8}{x}+17\right)
    \ln\left(\frac{1+r}{1-r}\right)\right)\right],
\end{eqnarray}
where $x\equiv s/M_{\Phi_{1}}$, $r=\sqrt{1-4/x}$, and \(g_{2}\) and
\(g_{Y}\) are the \(SU(2)\) and \(U(1)_{Y}\) coupling constants. The
calculation can also be performed for the non-supersymmetric case,
where the result is
\begin{eqnarray}
  \hat{\sigma}_{\mathit{SM}}^{\mathit{tot}}&=&
  \frac{1}{96\pi}\left[
  g_{Y}^{2}g_{2}^{2}\left(\left(36+\frac{144}{x}\right)r+
    144\left(\frac{2}{x^2}+\frac{1}{x}\right)
    \ln\left(\frac{1+r}{1-r}\right)\right)\right.\nonumber\\&&+
  3g_{2}^{4}\left(\left(39+\frac{196}{x}\right)r-
    144\left(\frac{2}{x^2}+\frac{3}{x}\right)
    \ln\left(\frac{1+r}{1-r}\right)\right)\nonumber\\&&+\left.
  g_{Y}^{4}\left(\left(53-\frac{116}{x}\right)r-
    \left(\frac{-96}{x^2}+\frac{48}{x}\right)
    \ln\left(\frac{1+r}{1-r}\right)\right)\right].
\end{eqnarray}
The effect of such second order annihilation processes on the
parameter space of Dirac leptogenesis is shown in
figure~\ref{fig:SecondOrderFX}, where, for comparison, we show two
sets of leptogenesis exclusion contours: one representing no
second-order processes and one representing annihilation in a
supersymmetric model.  It is evident from this graph that second
order processes do indeed lower the upper exclusion contour, though
the effect is not a dramatic one.

\end{document}